\documentclass[12pt,preprint]{aastex}

\shorttitle{Synchrotron Blob Model of Flares from Sgr A$^*$}
\shortauthors{Kusunose, \& Takahara}

\begin{document}

\title{Synchrotron Blob Model of Infrared 
and X-ray Flares from Sagittarius A$^*$}

\author{Masaaki Kusunose}
\affil{Department of Physics, School of Science and Technology,
Kwansei Gakuin University, Sanda 669-1337, Japan}
\email{kusunose@kwansei.ac.jp}

\and

\author{Fumio Takahara}
\affil{Department of Earth and Space Science,
Graduate School of Science, Osaka University,
Toyonaka 560-0043, Japan}
\email{takahara@vega.ess.sci.osaka-u.ac.jp}

\begin{abstract}
Sagittarius A$^*$ in the Galactic center harbors a supermassive black hole
and exhibits various active phenomena.
Besides quiescent emission in radio and submillimeter radiation,  
flares in the near infrared (NIR) and X-ray bands
are observed to occur frequently.  
We study a time-dependent model of the flares,
assuming that the emission is from a blob ejected from the central object.
Electrons obeying a power law with the exponential cutoff are assumed to be 
injected in the blob for a limited time interval.  
The flare data of 2007 April 4 were used to determine the values of model parameters.
The spectral energy distribution of flare emission is explained by nonthermal synchrotron
radiation in the NIR and X-ray bands. 
The model light curves suggest that electron acceleration is 
still underway during the rising phase of the flares.
GeV $\gamma$-rays are also emitted
by synchrotron self-Compton scattering, although its luminosity is not strictly
constrained by the current model.
If the GeV emission is faint, the plasma blob is dominated by the magnetic 
energy density over the electron kinetic energy density.
Observations in the GeV band will clarify the origin of the blob.
\end{abstract}


\section{Introduction}\label{sec:intro}

Various observations have confirmed that the Galactic center,
Sagittarius A$^*$ (Sgr A$^*$), contains a super massive black hole of mass
$\sim 4 \times 10^6 M_\sun$ \citep[e.g.,][]{ghez08,gill09a,gill09b}.
It has been found that Sgr A$^*$ emits radiation from radio through X-rays 
\citep[see][ for review]{melfal01,mel07} and even TeV $\gamma$-rays
\citep{tsu04,kos04,aha04,alb06}.
The bolometric luminosity of Sgr A$^*$ is $\sim 10^{36}$ erg s$^{-1}$
and the emission is dominated by radio.
The spectral flux sharply falls off above $\nu \sim 10^{12}$ Hz
(``the submillimeter bump'') \citep{zylka95,falcke98}.
In the quiescent state, X-rays at low luminosities were also observed 
by {\it Chandra}, which obtained the X-ray luminosity of
$L_X \approx 2.4 \times 10^{33}$ erg s$^{-1}$ in the 2 -- 10 keV band \citep{bag03}.
The quiescent state of Sgr A$^*$ can be described by a radiatively 
inefficient accretion flow (RIAF) model \citep[][]{yuan03} or a jet model \citep{fm00}. 
These models ascribe the submillimeter bump to synchrotron radiation by
thermal electrons in magnetic fields of $\sim 30$ G. 
The X-ray emission in the quiescent state is explained either by
bremsstrahlung or by inverse Compton scattering 
in both RIAF \citep{yuan03} and the jet models \citep{fm00}.

Strong flares occur frequently in the X-ray band \citep{bag01}
and near-infrared (NIR) band \citep{gen03}.
The detailed properties of the flares of NIR and X-rays are reviewed in \cite{de09}.
X-ray and NIR flares occur simultaneously, with no significant delay
\citep{eckart04,yuz06a}.
X-ray flares are always accompanied by NIR flares,
although NIR flares may occur without an associated X-ray flare
\citep[e.g.,][]{horn07}.  It should be noted, however, that
this might be instrumental, because the X-ray background in the Galactic center
is proportionally larger than the NIR background and, thus, 
only bright X-ray flares are detected.
The duration of NIR flare in the $L'$ band (3.80 $\mu$m) observed in 2007 April 4 was
about 100 minutes \citep{de09}.
Simultaneously, {\it XMM-Newton} observed the flare in the X-ray band 
and the X-ray flare lasted about 60 minutes \citep{por08}.
The $L'$-band light curve of the 2007 flare has sub-structural variations 
on a timescale of $\sim 20$ minutes.  
\cite{de09} attribute the NIR substructure to the fluctuation of the magnetic field.

Observations of polarization of the IR flares 
are broadly consistent with that the IR flares are synchrotron origin
\citep[e.g.,][]{eckart06,marrone08,de09,yuz09}.
On the other hand, the emission mechanisms of the X-ray flares are still debated;
upscattering of submillimeter photons \citep{mar01, yuz06a},
synchrotron self-Compton (SSC) scattering  \citep{yuan03,eckart04,eckart06,eckart06b,sabha10},
synchrotron emission from high energy electrons
\citep{yuan03,yuan04,de09},
or inverse Compton scattering of NIR photons by $\sim 10$ MeV electrons
responsible for the quiescent radio-millimeter emission \citep{yuz09}.
An orbiting hot spot model is also proposed \citep{bl05,bl06}.
This model requires the synchrotron cooling time longer than the orbital period
to explain the light curve of NIR.
However, the cooling time of X-ray emitting electrons is much shorter than the NIR
emitting electrons.  This makes the hot spot model inappropriate to
the X-ray flare emission.

\cite{de09} explored the emission mechanisms of the 2007 IR/X-ray flare by
various synchrotron and inverse Compton emission models.
According to their model, SSC models need a large magnetic field such as
6000 G and a very small size of the flare emission region, 
e.g., $\sim 0.0013 R_\mathrm{S}$,
where $R_\mathrm{S}$ is the Schwarzschild radius.
The magnetic field is too large compared with the inferred value during
the quiescent state, i.e., 10 - 30 G,
and the size is too small to account for the flare timescale of 100 minutes.
On the other hand, the flare is well explained by ``powerlawcool'' model, 
in which electrons with a power law with a cooling break emit IR and X-rays 
by synchrotron radiation.  
Recently, \cite{sabha10} explored SSC models that include parameter regions 
not considered by \cite{de09}.  Their model also requires a smaller source component
size to give a sufficient inverse Compton scattering efficiency,
so that multi-components are needed to obtain a broad time profile of a flare.
\cite{marrone08} also constructed a SSC model of X-ray emission for a flare
observed in 2006 July 17.
The source size of their model is $\sim R_\mathrm{S}$ and the strength
of magnetic fields is 1.5 G.
The source size and the magnetic fields are smaller than those of our model 
by a factor $\sim 10$ as shown in Section \ref{sec:numresults}.

Because the origin of X-ray emission is still unclear as described above,
in this paper we present a model that NIR and X-ray flares can be produced by 
synchrotron radiation from a single plasma blob.
The simultaneity of NIR and X-ray flares favors a model of emission from a single region.
The blob might be ejected from the region near the central black hole.
As shown below, the size of the blob of our model is about 10 $R_\mathrm{S}$ 
and magnetic field is about 20 G.
These values are consistent with the physical parameters of the accretion flow
around the central black hole.
In this model the flare duration is determined by the injection timescale of
nonthermal electrons.
The decay of the flares is owing to the radiative cooling and particle escape
after turnoff of the injection of nonthermal electrons.  
In particular, we focus on the time-dependent behavior of the flare models.
Because of various timescales such as cooling, injection, escape, etc.,
time-dependent simulations are necessary to investigate the flare properties in detail.
It should be noted that most of the previous theoretical work on the flare emission 
has been restricted to the steady state models.

Although we do not consider adiabatic expansion in this paper,
there are phenomena suggesting the effects of adiabatic expansion. 
For example, \cite{horn07} found that their NIR measurements are
consistent with a constant spectral index during a flare.
\cite{marrone08} and \cite{yuz09}, on the other hand,
found that millimeter and submillimeter peaks delay the peak of 
NIR/X-ray flares.
According to their models this delay is consistent with 
the adiabatic expansion of a self-absorbed source.
The estimated expansion speed is much less than $0.1 c$.

Our assumptions on the emission region are described in Section \ref{sec:model}.
Numerical results are given in Section \ref{sec:numresults}.
A summary of our results and discussion are given in Section \ref{sec:sum}.
Throughout this paper, we assume 8 kpc as a Galactic center distance 
\citep{eis03} and $4 \times 10^6 M_\sun$ as the central black hole
mass \citep{ghez08,gill09a,gill09b}.


\section{Blob  Model} \label{sec:model}

We first assume that the emission region of the flares is a blob ejected from 
the central region around the black hole, e.g., from accretion flow
\citep{wan00,wk02,yuan09}.
For simplicity of numerical calculations,
we assume that the blob is a uniform sphere with radius $R$ and magnetic field $B$. 
While RIAF or a jet supplies the quiescent component of the spectral energy
distribution (SED) of radiation, mainly in radio, 
the temporal ejection of a blob produces flare emission.
Since the blob is launched from the inner edge or near the inner edge of the accretion
flow, the blob speed might be at least mildly relativistic.
However, we assume that the blob moves
at nonrelativistic speed, so that the relativistic effects such as beaming are neglected.
Because relativistic jets have not been observed from the region near Sgr A$^*$,
it is possible to neglect the effects of the motion of the blob 
as a first step of investigation.

In our simulations the injection of nonthermal electrons in the blob triggers flares.
Shock acceleration in the blob or magnetic reconnection can supply such nonthermal 
electrons.
We solve the kinetic equations of electrons and photons as 
in \cite{kus00} and \cite{lk00}.
We include the injection of nonthermal electrons, radiative cooling (synchrotron and
inverse Compton scattering), and escape from the blob.
The injection rate per unit volume and unit interval of $\gamma$,
where $\gamma$ is the electron Lorentz factor, is given by
\begin{equation}
q(\gamma) =
\cases{
0 , & \mbox{if}  $\gamma < \gamma_\mathrm{min}$ , \cr
K \gamma^{-p} \exp(-\gamma/\gamma_\mathrm{max}) ,
 & \mbox{if} $\gamma \geq \gamma_\mathrm{min}$ , \cr
}
\end{equation}
where $K$, $p$, $\gamma_\mathrm{min}$, and
$\gamma_\mathrm{max}$ are parameters.  
The value of $\gamma_\mathrm{max}$ can be a function of time,
which allows different shapes of light curves.
In our model described below, we assume that $\gamma_\mathrm{max}$ increases
linearly with time to fit the flare light curves of 2007 April 4.
The normalization factor, $K$, is determined by assigning the value of
the injection rate per unit volume, i.e., 
\begin{equation}
q_\mathrm{inj} = \int_{\gamma_\mathrm{min}}^\infty q(\gamma) d \gamma .
\end{equation}
A time-dependent $q_\mathrm{inj}$ model is also considered 
(see Section \ref{sec:numresults}).
The injection duration of nonthermal electrons, $t_\mathrm{inj}$, 
and the escape timescale of electrons from the blob, $t_\mathrm{esc}$,
are also parameters.
The emission mechanisms are synchrotron radiation by the nonthermal electrons
and inverse Compton scattering of the synchrotron photons by the same electrons
(SSC).

In this paper we do not consider Comptonization of external soft photons such as
radiation from accretion flow.  We discuss the validity of this assumption in
Section \ref{sec:sum}.
Furthermore, the adiabatic expansion of an initially optically thick blob 
is excluded, because there is no significant delay or asymmetry in the longer 
wavelength emission relative to the peak of the X-ray flare \citep{de09}.

\section{Numerical Results}  \label{sec:numresults}

Numerical models are obtained for the parameters such as
$R$, $B$, $q_\mathrm{inj}$, $p$, $\gamma_\mathrm{min}$,
$\gamma_\mathrm{max}$, $t_\mathrm{inj}$, and $t_\mathrm{esc}$.
For a black hole mass $M_\mathrm{BH} = 4 \times 10^6 M_\sun$,
the Schwarzschild radius is 
$R_\mathrm{S} = 2 GM_\mathrm{BH}/c^2 = 1.2 \times 10^{12}$ cm.
The light crossing time of this size is  
$R_\mathrm{S}/c \sim 40$ s.
The particle escape occurs by advection, so that we
set $t_\mathrm{esc} \sim $ several $R/c$.
In our model the injection duration is strongly related to the flare timescale.
The values of parameters are determined by the SED of IR and X-ray flares
and light curves.
For this purpose, we use the data of the 2007 IR/X-ray flare given
in \cite{por08} and \cite{de09}.

\subsection{Light Curves}

In our model, the emission in both IR and X-ray bands is produced
by synchrotron radiation.
The maximum value of $\nu F_\nu$ of synchrotron emission appears 
at $\sim 10^{16}$ Hz, which is in between the IR and X-rays bands,
where $F_\nu$ is the energy flux per unit frequency.
The luminosity of X-rays is sensitive to the values of $\gamma_\mathrm{max}$
and $t_\mathrm{inj}$.  This is because X-rays are emitted by the electrons 
in the high energy end of the electron spectrum and the cooling time of 
those electrons is short.
As soon as the injection rate of nonthermal electrons decreases,
the X-ray luminosity declines.
Because IR emitting electrons have longer cooling time than X-ray emitting electrons, 
the IR luminosity is more dependent on $t_\mathrm{esc}$ than the X-ray luminosity.
If $t_\mathrm{esc}$ is long enough, the X-ray emitting electrons begin to
pile up as lower energy electrons that emit IR.

\begin{figure}
\includegraphics[angle=0,scale=.65]{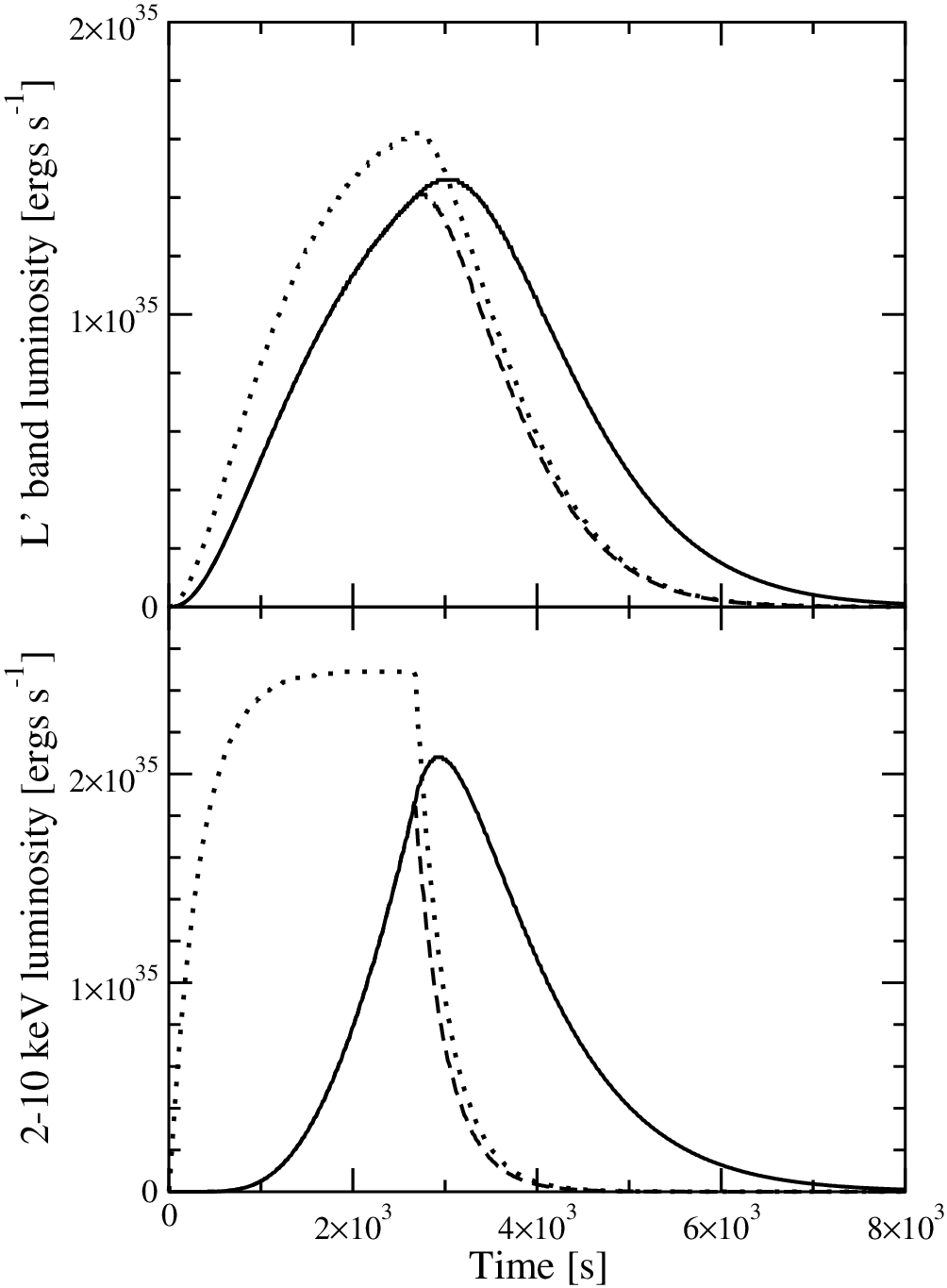}

\caption{
Light curves of the $L'$ band ($3.8 \mu$m) and the 2 -- 10 keV band
for various models.  A model with a fixed value of 
$\gamma_\mathrm{max} = 5 \times 10^4$ is shown by dotted lines.
The dashed lines are for a model with a time dependent $\gamma_\mathrm{max}$.
Here we assume that $\gamma_\mathrm{max}$ increases linearly from 500 
to $5 \times 10^4$ during $t = 0$ and $t_\mathrm{inj}$,
where $t_\mathrm{inj} = 2.7 \times 10^{3}$ s.
Both models assume that the injection is at a constant rate 
and stops at $t = t_\mathrm{inj}$.
A model with a time dependent injection is shown by solid lines.
This model assumes that $q_\mathrm{inj}$ is constant for $t < t_\mathrm{inj}$ 
and decreases as 
$q_\mathrm{inj} \propto \exp[-(t-t_\mathrm{inj})/(0.25 t_\mathrm{inj})]$
for $t > t_\mathrm{inj}$.  The value of $\gamma_\mathrm{max}$ increases
linearly as for the model shown by the dashed lines.
For all these models, $R = 10^{13}$ cm, $B = 20$ G, $\gamma_\mathrm{min} = 2$,
$p = 1.3$, $t_\mathrm{inj} = 8 R/c$, and $t_\mathrm{esc} = 5 R/c$ are used.
\label{fig:typical-lightcurves}}
\end{figure}

We first examine light curves of the IR and X-ray bands for various models.
For models in this subsection, 
we assume that $R = 10^{13}$ cm, $B = 20$ G, $\gamma_\mathrm{min} = 2$,
$p = 1.3$, $t_\mathrm{inj} = 8 R/c$, and $t_\mathrm{esc} = 5 R/c$.
The blob is initially empty at $t = 0$ and nonthermal electrons are injected for $t > 0$.
In Figure \ref{fig:typical-lightcurves}, 
we present light curves in the $L'$ (3.8 $\mu$m) and 2 -- 10 keV bands for different
models of $\gamma_\mathrm{max}$ and $q_\mathrm{inj}$.
When $\gamma_\mathrm{max}$ is fixed, high-energy electrons are injected instantly
at $t = 0$.
Then the light curves rise steeply (dotted lines).
It is found for this model that the light curve of X-rays increases faster
than that of IR, which is not in agreement with the flare of 2007 April 4.
To obtain slower rise of the light curves, the time dependent $\gamma_\mathrm{max}$ 
is assumed as shown by the dashed lines.
Here we assumed that $\gamma_\mathrm{max}$ increases linearly with time
from 500 to $5 \times 10^4$ during $t = 0$ and $t_\mathrm{inj}$.
For $t > t_\mathrm{inj}$ both models (the dotted and dashed lines) 
assume $q_\mathrm{inj} = 0$. 
By fitting model light curves with the data of the flares,
we found that the above models have shorter decline timescales than the observed one.
We then assume that the injection rate decreases gradually after $t = t_\mathrm{inj}$.
As an example, we use a model with 
$q(\gamma) \propto \exp[-(t-t_\mathrm{inj})/(\xi t_\mathrm{inj})]$
for $t > t_\mathrm{inj}$.
In Figure \ref{fig:typical-lightcurves},
a model with $\xi = 0.25$ is shown by the solid lines (model A).
In this model $\gamma_\mathrm{max}$ is constant for $t > t_\mathrm{inj}$.

\begin{figure}
\includegraphics[angle=0,scale=.7]{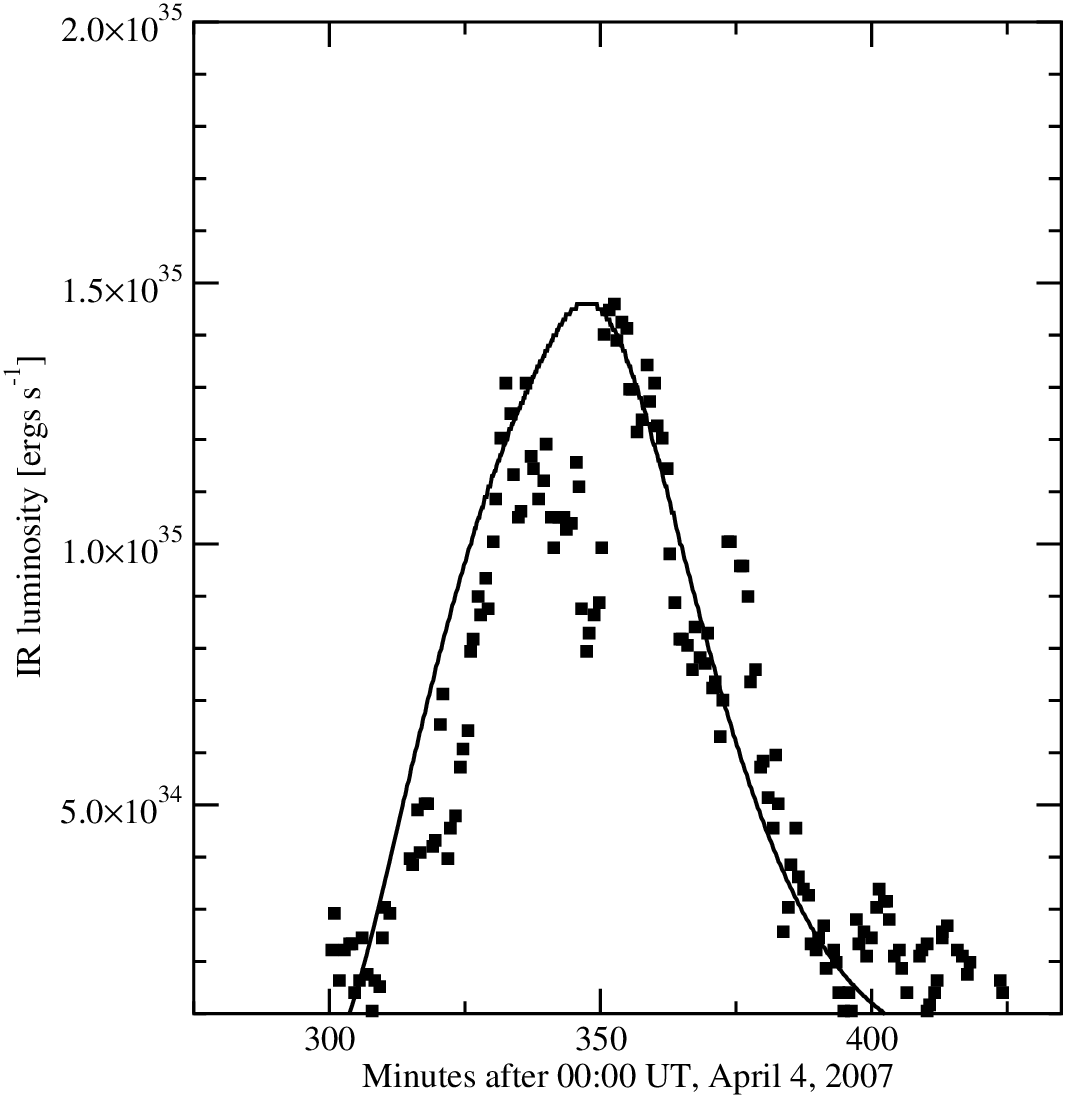}
\caption{
Light curves of the $L'$ band.
The flare data of the $L'$ band (filled squares) are from \cite{de09}.
The solid line depicts model A.  The value of $\gamma_\mathrm{max}$ increases 
linearly with time from 500 up to $5 \times 10^4$.
In Figures 2 and 3, the peak time of model light curves are shifted
so that the peak time coincides with the observed peak time.
\label{fig:lightcurves-fit-IR}
}
\end{figure}

\begin{figure}
\includegraphics[angle=0,scale=.7]{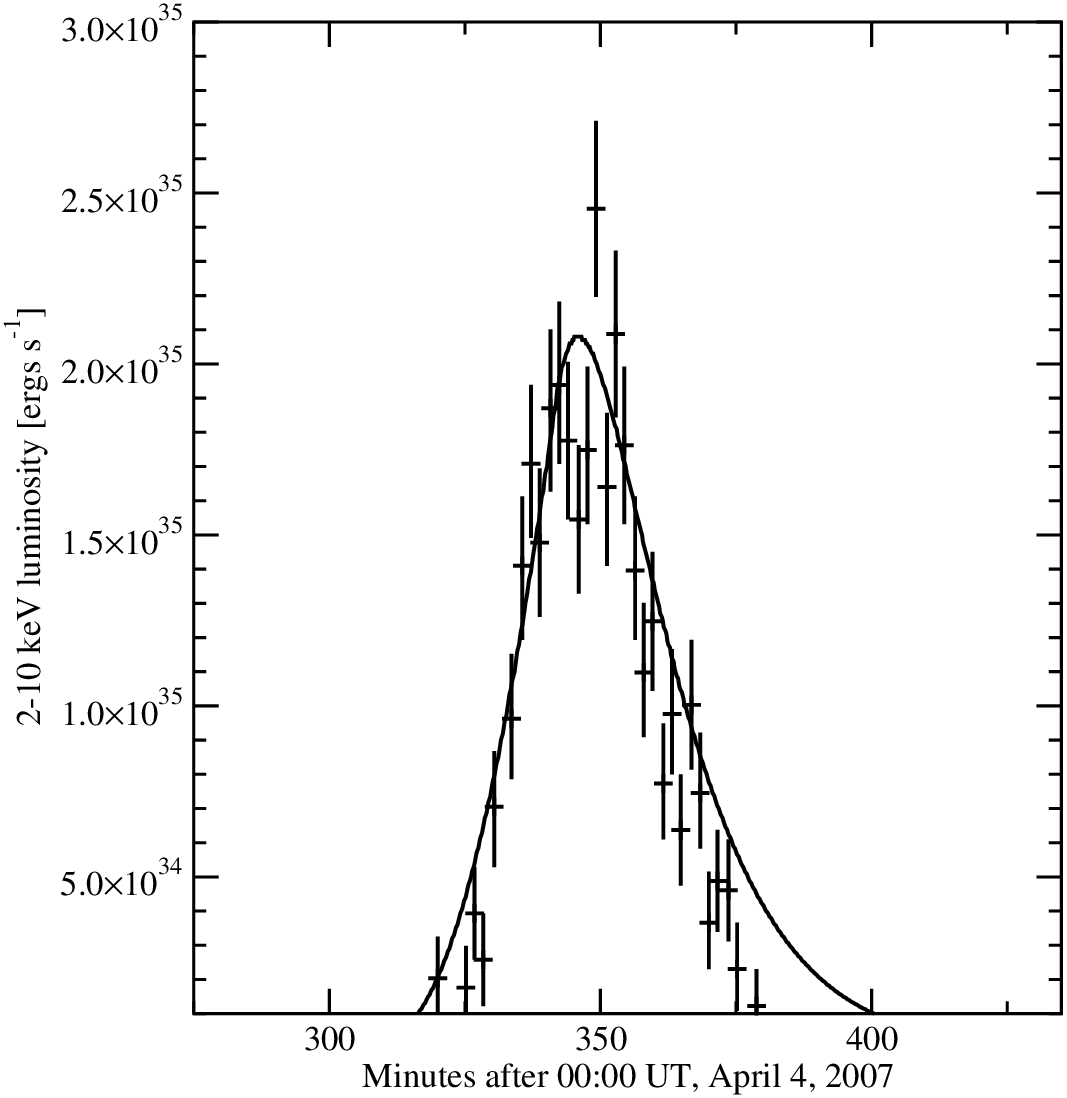}
\caption{
Light curves of the 2 -- 10 keV band.
The flare data of the X-ray band (pluses) are from \cite{por08}. 
Model A is shown by a solid line.
\label{fig:lightcurves-fit-X}
}
\end{figure}

In Figures \ref{fig:lightcurves-fit-IR} and \ref{fig:lightcurves-fit-X},
we compare model light curves to the observed data for 2007 April 4.
Here the model light curves are the same as shown by the solid lines in Figure 
\ref{fig:typical-lightcurves}, i.e., model A.  
The details of the model parameters are given in Table \ref{table:param-values1}.
The size of the blob $R$ is $10^{13}$ cm and 
this corresponds to the light crossing time of 334 s.
In this figure, the peak time of the luminosity of the model is shifted
so that the model curves coincide with the observed light curves.
We assumed the linear dependence of $\gamma_\mathrm{max}$ on time 
for $t < t_\mathrm{inj}$ 
and the gradual decline of the injection rate for $t > t_\mathrm{inj}$.
The fact that IR and X-rays attain the peak flux almost simultaneously \citep{de09}
implies that the value of $\gamma_\mathrm{max}$ increases during the rising phase
of the light curves, i.e., 
the acceleration of higher energy electrons is still underway 
in the early phase of the flare.

\begin{figure}
\includegraphics[angle=0,scale=.7]{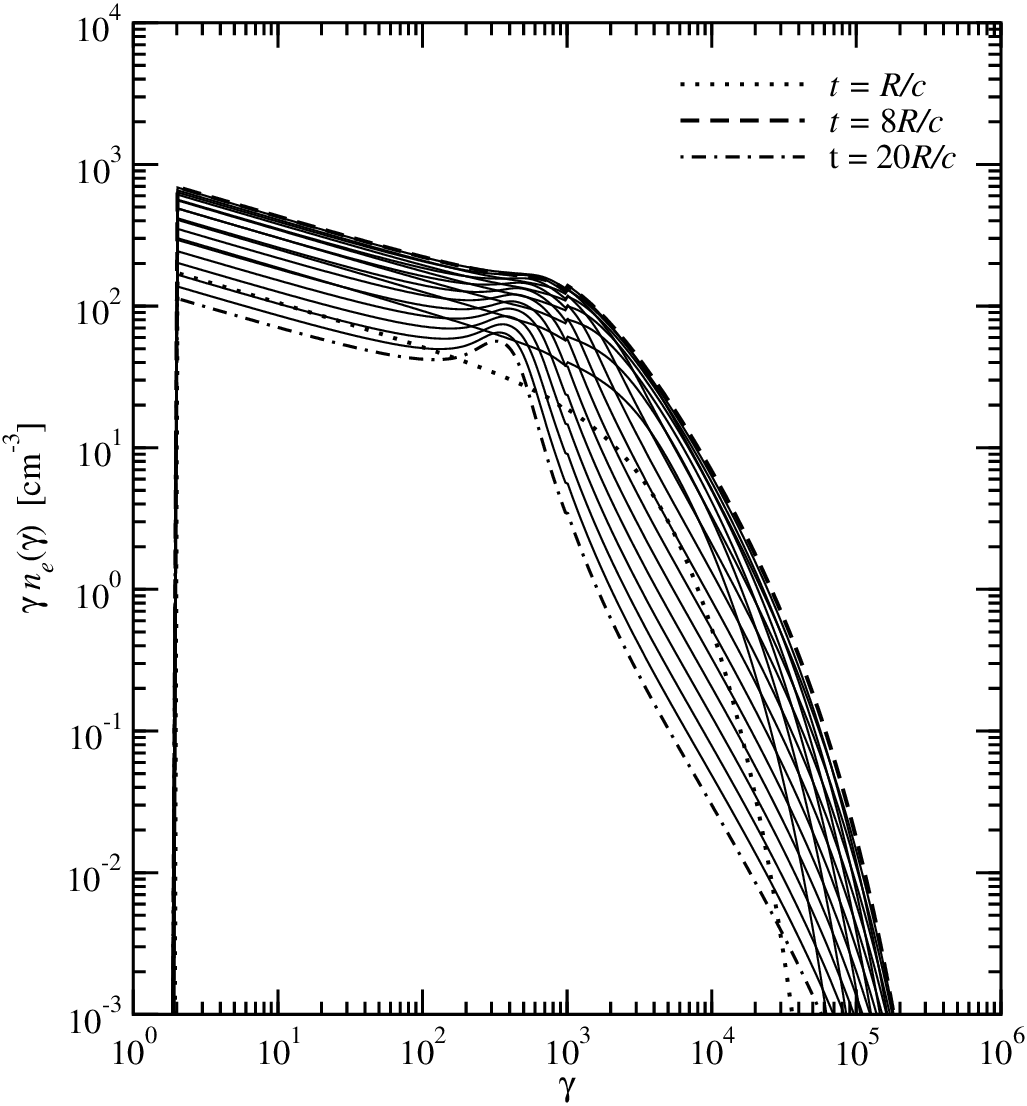}
\caption{Time evolution of the electron spectrum of model A.
Here, $n_e(\gamma)$ is the electron number density per unit $\gamma$.
The spectra are shown for every $R/c$.
After $t = 8R/c$, the injection rate decreases exponentially.
\label{fig:electron-time}}
\end{figure}

The cooling time of X-ray emitting electrons is $\sim 11$ s and much shorter
than the escape time $t_\mathrm{esc} \sim 1.7 \times 10^3$ s,
if 10 keV of X-rays are assumed.
On the other hand, NIR is emitted by electrons with $\gamma \sim 10^3$.
The synchrotron cooling time of those NIR emitting electrons is 
comparable with $t_\mathrm{esc}$, i.e., $t_\mathrm{cool} \sim 2 \times 10^3$ s.
Thus the cooling is effective in X-rays during the decay of the flare
and the decline time of the X-ray light curve is shorter than that of NIR.
The decline of the NIR light curve is regulated by both $t_\mathrm{cool}$ and
$t_\mathrm{esc}$.   As shown in Figure \ref{fig:electron-time}, 
a bump appears at $\gamma \sim 10^3$ in the electron spectrum.
This is because of balance between radiative cooling and escape.

\subsection{Spectral Energy Distribution}

\begin{figure}
\includegraphics[angle=0,scale=.7]{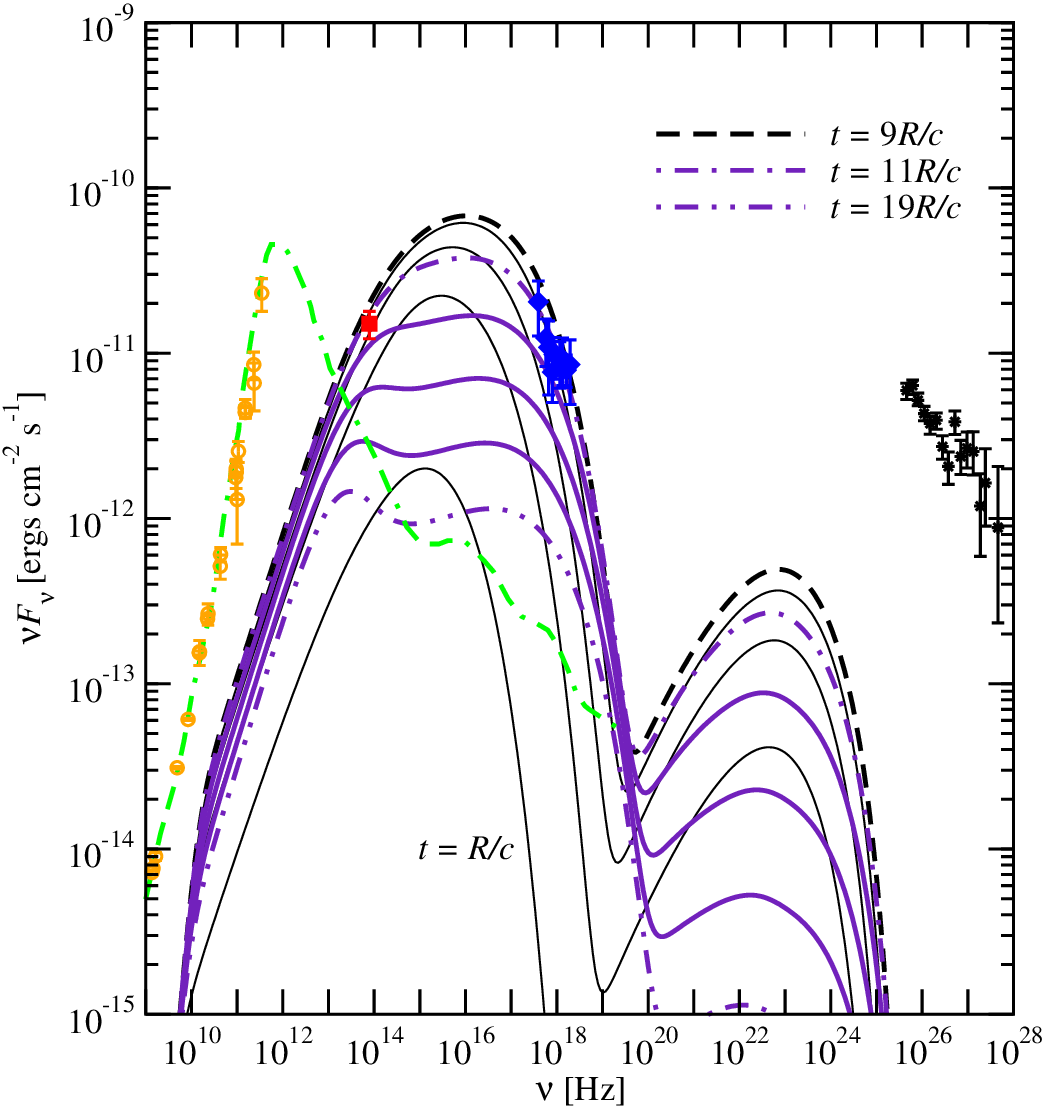}
\caption{
Time evolution of the SED of model A.  The SEDs are shown for
$t= R/c$ -- $19R/c$ at every $2 R/c$.
SEDs in the rising phase are shown by thin lines and those in the 
decaying phase by thick lines.
Radio to submillimeter measurements are for the quiescent state
\citep{mar01,zhao03} (open circles).
The flaring state in NIR (filled square) is taken from \cite{de09}.
The X-ray flare data (filled diamonds) are from \cite{por08}.  
TeV emission \citep{aha04} is also shown by asterisks.
The quiescent state RIAF model by \cite{yuan03} is shown by a dash-dash-dotted
line for comparison.
\label{fig:SED-time}}
\end{figure}

We next show the evolution of the flare SED of model A 
and compare the model to the observations in Figure \ref{fig:SED-time}.
The quiescent model by \cite{yuan03} is also shown for comparison.
The maximum flare luminosity is $5.0 \times 10^{37}$ erg s$^{-1}$
at $t = 9 R/c$.
This luminosity is much higher than the submillimeter luminosity 
in the quiescent state,
but this flare does not strongly affect the spectrum in the submillimeter wavelength.
During the rising phase, synchrotron emission becomes luminous
and attains the peak at $t \sim 8 R/c$.
The peak frequency increases because $\gamma_\mathrm{max}$ increases
for $t < t_\mathrm{inj}$.
While the emission decays, the spectrum in the IR and X-ray bands
become flatter and a dip appears at $\sim 10^{15}$ Hz.

It is noted that the SSC component appears in the GeV region, whose intensity might be
too small to be observed.  However, other sets of parameter values
allow a stronger GeV flux as shown below.

\begin{figure}
\includegraphics[angle=0,scale=.7]{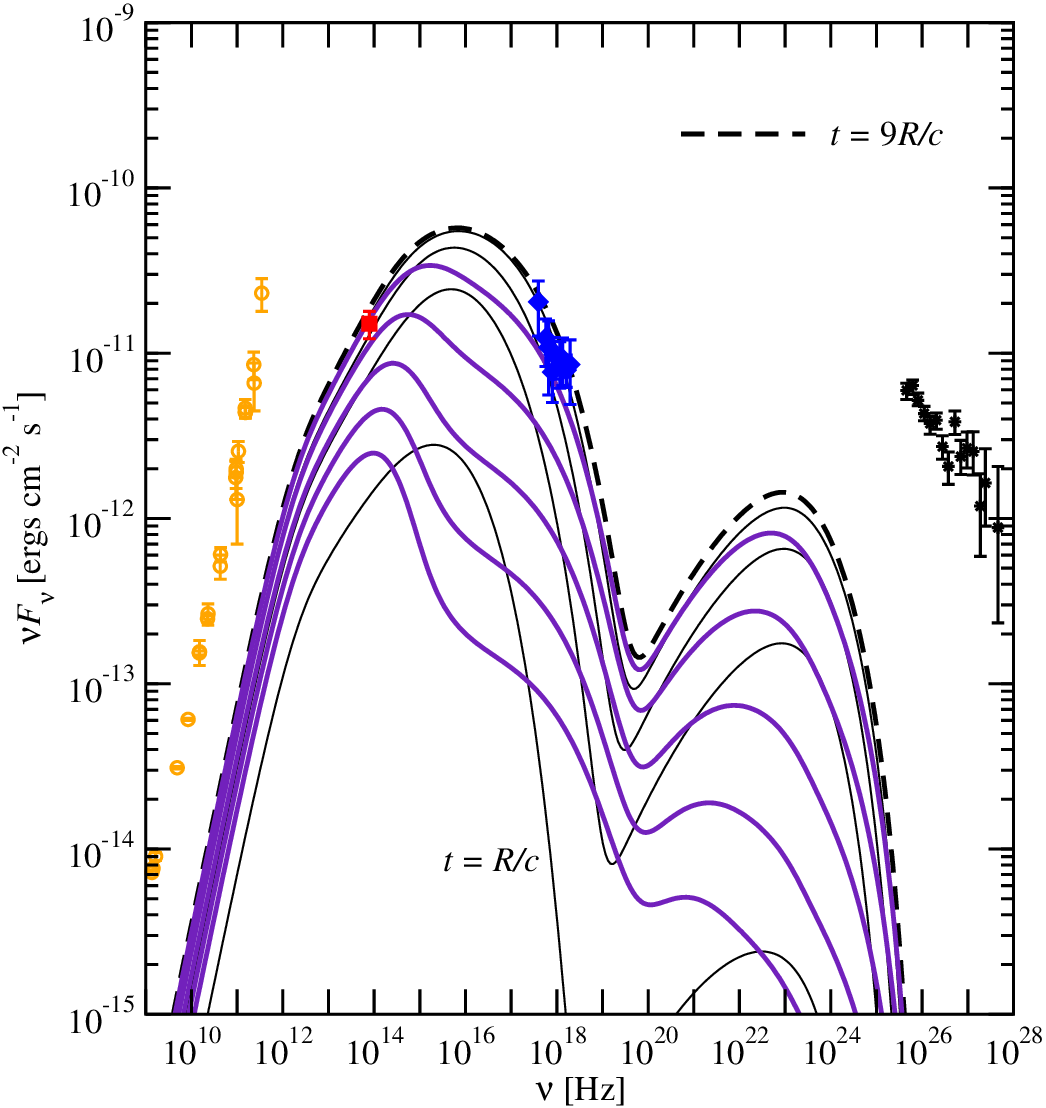}
\caption{Time evolution of the SED of model B with $p = 1.8$.
Here, $\gamma_\mathrm{min} = 2 \times 10^2$ is assumed.
The value of $\gamma_\mathrm{max}$ is increased from 
$1.3 \times 10^3$ up to $1.3 \times 10^5$ linearly with time 
during $t = 0$ and $t_\mathrm{inj}$.
SEDs in the rising phase are shown by thin lines and those in the 
decaying phase by thick lines.
\label{fig:sed-evolve-p-diff}
}
\end{figure}

Since the number of parameters is large and observed data are few, 
there is uncertainty in determining the parameter values to fit the data.
We show another model (model B) with $p = 1.8$ in Figure \ref{fig:sed-evolve-p-diff}.
When $p$ has a larger value, the value of $\gamma_\mathrm{min}$ should be larger for
the radio flare not to exceed the quiescent radio emission.
When the value of $p$ is larger and the value of $\gamma_\mathrm{min}$ is fixed, 
there are more electrons with lower energy for the same IR and X-ray luminosities
as the observed data.
This constraint on $\gamma_\mathrm{min}$ may be not strong,
if flares in radio occur simultaneously.

\begin{figure}
\includegraphics[angle=0,scale=.7]{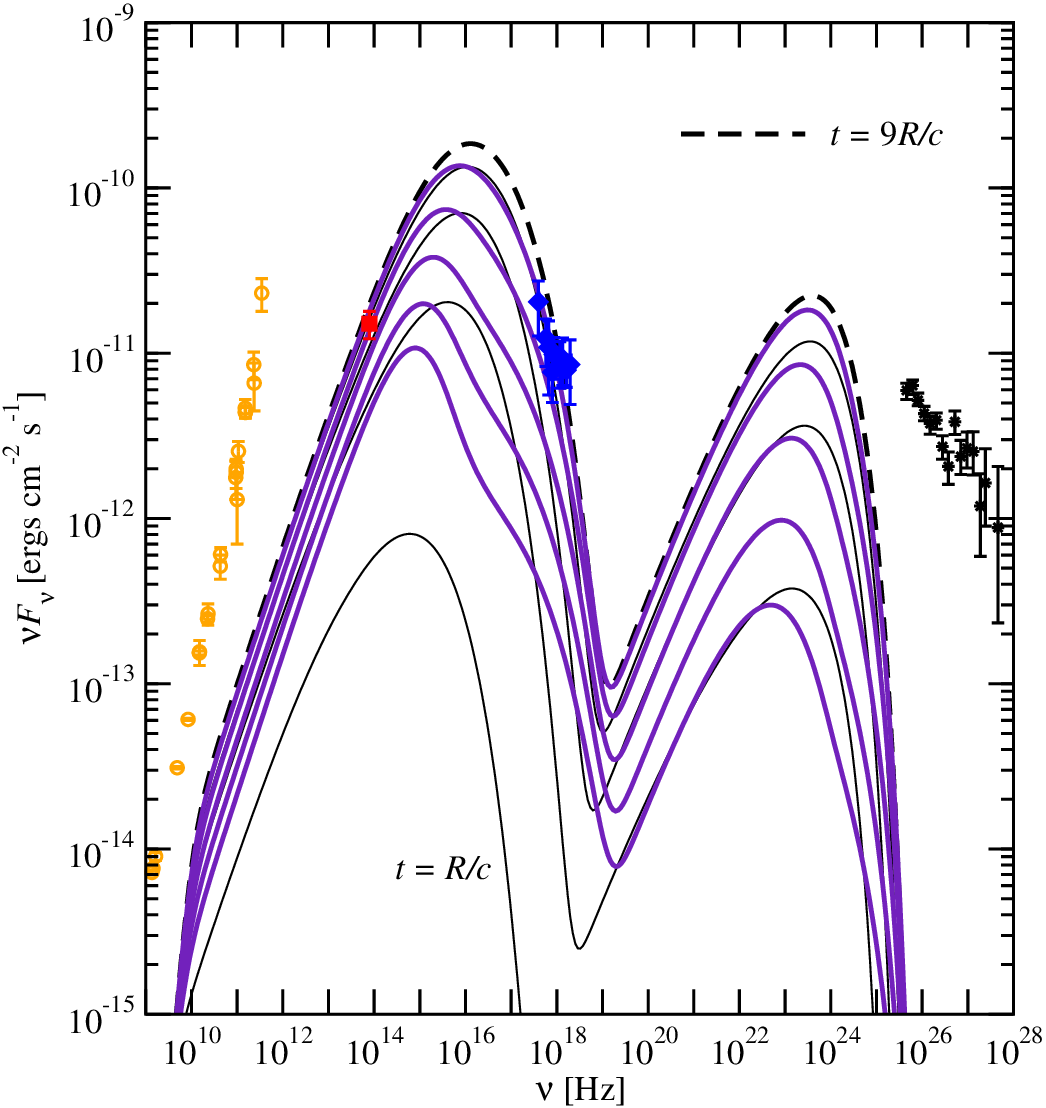}
\caption{Time evolution of the SED with a luminous SSC component (model C).
Here, $B = 5$ G and $q_\mathrm{inj} = 7.5$ are assumed.
SEDs in the rising phase are shown by thin lines and those in the 
decaying phase by thick lines.
\label{fig:sed-evolve-SSC}
}
\end{figure}

As shown in Figure \ref{fig:SED-time}, our model predicts the emission in the GeV band,
which is produced by inverse Compton scattering of IR-optical photons, i.e., SSC process.
The luminosity in the GeV band is not strongly constrained in our model.
In Figure \ref{fig:sed-evolve-SSC}, we show a model with a larger luminosity 
in the GeV band (model C).
A smaller value of the magnetic field and a higher injection rate produce
a luminous SSC component.
The difference of SEDs in Figures \ref{fig:SED-time} and \ref{fig:sed-evolve-SSC} 
also appears in the IR band of the decaying phase.
In Figure \ref{fig:SED-time} the IR spectrum becomes flat after $t \sim 12 R/c$, 
while that in Figure \ref{fig:sed-evolve-SSC} keeps hard spectral shape.

\subsection{Electron Spectrum and Energetics}

The time evolution of the electron spectrum of model A is shown 
in Figure \ref{fig:electron-time}.
The number density of electrons increases for $t < t_\mathrm{inj}$ 
because of injection.  
For $t > t_\mathrm{inj}$, the injection rate decreases exponentially with time.
The higher energy electrons decrease because of radiative cooling.
Also on timescale $t_\mathrm{esc}$, electrons escape from the blob.

\begin{figure}
\includegraphics[angle=0,scale=.7]{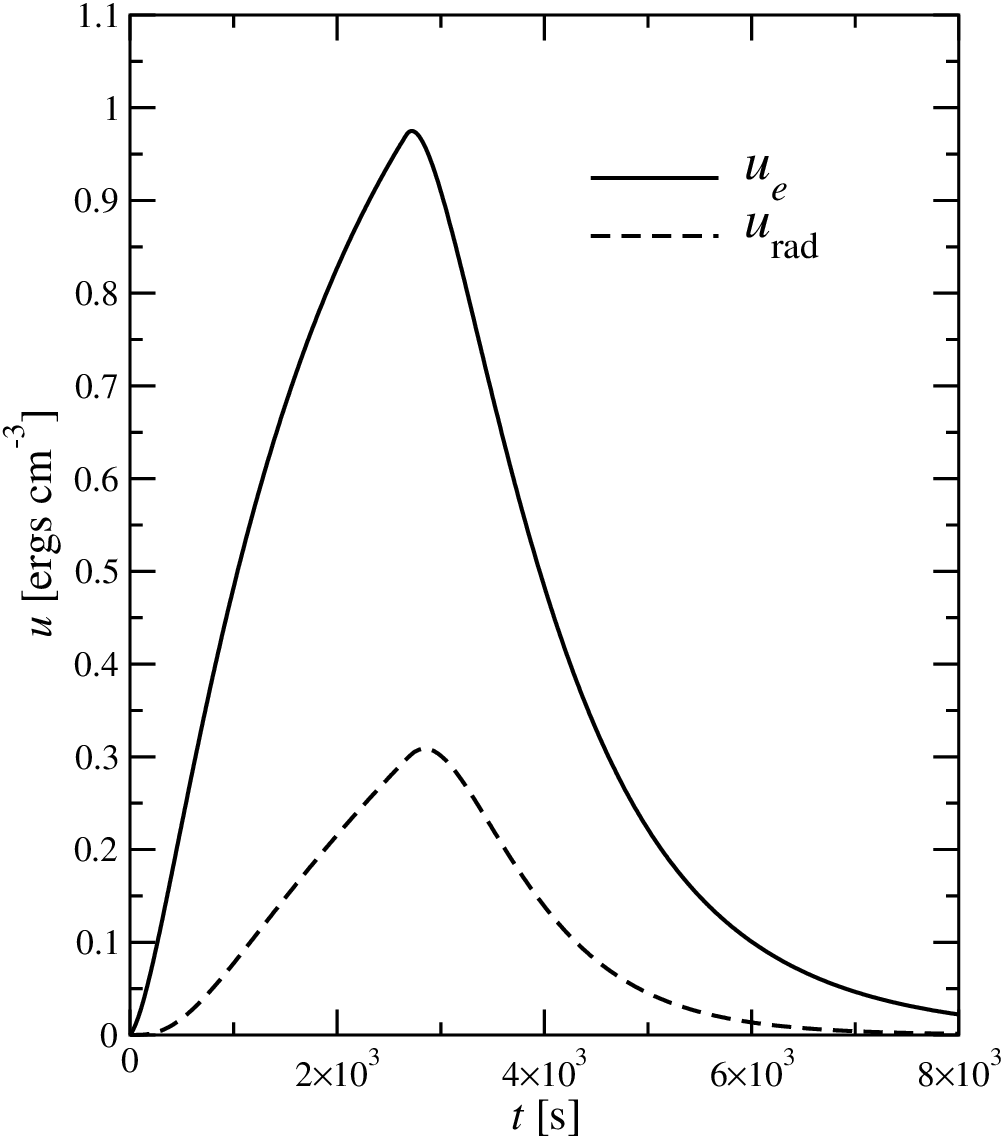}
\caption{Time evolution of the energy densities of electrons (solid) 
and radiation (dashed).
This model assumes that the magnetic energy density is constant, i.e., 
$B^2/(8 \pi) = 15.9$ erg cm$^{-3}$ with $B = 20$ G.
\label{fig:energy}
}
\end{figure}

The time evolution of the energy densities of 
the radiation $u_\mathrm{rad}$ and the kinetic energy of electrons $u_e$ 
is shown in Figure \ref{fig:energy} for model A.
In our model the magnetic field is assumed to be constant.
The energy content is dominated by the magnetic energy density $u_B$, i.e.,
$u_B/u_e \sim 16$ at $t \sim 8 R/c$.
This is in contrast with blazar models, where the kinetic energy of 
nonthermal electrons/positrons dominates over the magnetic energy \citep{ktk02}.
Blazars are relativistic jets aligned with the line of sight and 
the relativistic beaming effect enhances the luminosity,
while our model of Sgr A$^*$ does not have relativistic motion.
It should be also noted that blazars have a luminous SSC or external Compton
component in the $\gamma$-ray band, while model A has only a weak SSC 
component.  When the SSC component is more luminous as in model C,
the electron energy dominates over the magnetic energy, i.e.,
$u_B/u_e \sim 0.1$ at $t \sim 8 R/c$.

\section{Summary and Discussion} \label{sec:sum}

We calculated the SEDs and light curves from a blob as a model of
the IR and X-ray flares observed from Sgr A$^*$.
Because we performed time-dependent calculations, 
it was possible to study 
the evolution of radiation and electron spectra in detail,
which is different from most of previous work.
Our model assumes that IR and X-rays are emitted by synchrotron radiation 
of nonthermal electrons injected in the blob.
This model explains the characteristics of 
2007 April 4 flare in IR/X-rays such as
(1) the simultaneity of the X-ray and IR flares,
(2) the steeper decline of the X-ray flare than the IR flare,
(3) the flare timescale of 100 minutes.
The flare timescale in our model is mainly determined by the injection time
of nonthermal electrons, which is related to particle acceleration mechanisms
and to be studied in future work.
To compare the light curves we assumed that the value of $\gamma_\mathrm{max}$ 
increases linearly with time.
If the value of $\gamma_\mathrm{max}$ is fixed during injection,
our model does not explain the simultaneity of the X-ray and IR flares.
The higher energy electrons responsible for X-ray emission are thus being 
accelerated in the rising phase of the flare.

We compared the SEDs of different models.
While model A has a faint SSC component in $\gamma$-rays and a larger value of $u_B/u_e$, 
model C has a bright SSC one and a smaller value of $u_B/u_e$.
If model C is applicable, we will observe $\gamma$-rays in the GeV band.
Future observations in the GeV band will constrain the model parameters
as well as the energetics in the flare regions.
Because the blob might be ejected from RIAF by magnetohydrodynamic processes
and nonthermal electrons are accelerated in the collisonless shock,
the observations of GeV $\gamma$-rays will clarify the origin of the blobs and 
the particle acceleration mechanisms.

The spectral index of NIR is almost constant during a flare
in model C, while the spectrum becomes flatter during the decay
in models A and B.  In model C, the cooling time is longer
than that of models A and B, because the magnetic field is weaker.
Then only escape time, which is independent of electron energy,
regulates the decay of the NIR flare.

Previous work mainly modeled the SED of flares by steady state models
and did not utilize the data of light curves.
Time-dependent models were recently published by \cite{sabha10}.
They assumed SSC for X-ray flares and combined adiabatic expansion to account for
light curves of 2007 April 4.  To fit the data six source components are needed.
While it is possible to attribute the IR substructure to those multi-components 
of sources,
it seems difficult to produce many emission regions coherently to cause flares.

It is known that IR flares sometimes occur without X-ray flares.
This might be owing to a small value of $\gamma_\mathrm{max}$ 
because of inefficient acceleration, rapid radiative cooling, or adiabatic cooling.  
If, for example, a blob is close to RIAF and external Compton scattering is
important, synchrotron X-ray flares are suppressed.
If this is the case, the effect of external Compton scattering 
appears as weak X-ray flares or hard spectral X-ray emission.
This is because when synchrotron X-ray flares are absent,
the kinetic energy of electrons with $\gamma \gtrsim 10^3$ are given to 
the RIAF photons with $\nu \sim 10^{12}$ Hz, yielding X-rays with $\nu \gtrsim 10^{18}$ Hz.
The intensity of these X-rays depends on the distance of the blob from RIAF.

We neglected the effect of the radiation from the RIAF in the flare emission spectra.
By inverse Compton scattering of submillimeter photons with $10^{12}$ Hz, 
electrons with $\gamma_\mathrm{max} \sim 10^4$ produces gamma-rays with $\sim 1$ MeV.  
If the submillimeter emission ($L \sim 3.85 \times 10^{35}$ erg s$^{-1}$) 
is received by the blob, the energy density of the submillimeter photons 
in the blob is $\sim 1.8 \times 10^{-2}$ erg cm$^{-3}$.
Here we assumed that the photon density decreases as the square of the distance,
that the distance of the blob from the submillimeter emission region is 1 AU,
and that the submillimeter emission is isotropic.
This energy density should be compared with that of synchrotron photons in the blob, 
$\sim 0.3$ erg cm$^{-3}$ at $t \sim 8 R/c$ (model A).  
Thus external Compton scattering of submillimeter photons is negligible,
if the distance is greater than about 1 AU.

After completion of this work, we noticed a paper by \cite{de10}.
They solved a kinetic equation of electrons and
obtained time-dependent flare spectra of synchrotron radiation:
Note that they did not calculate SSC spectra.
In their work flares are caused by the change in the injection rate and
magnetic fields, while we assumed the change in the injection rate and
$\gamma_\mathrm{max}$.



\clearpage


\clearpage
\begin{deluxetable}{ccccccccc}
\tablecaption{Parameters \label{table:param-values1}}
\tablewidth{0pt}
\tablehead{
\colhead{Model} & \colhead{$B$(G)} & \colhead{$p$} & 
\colhead{$\gamma_\mathrm{min}$} & \colhead{$\gamma_\mathrm{max}$} & 
$q_\mathrm{inj}$(cm$^{-3}$ s$^{-1}$) &
}
\startdata
A & 20 & 1.3 & 2 & $5 \times 10^4$ & 1.6  \\        
B & 10 & 1.8& $2 \times 10^2$ & $1.3 \times 10^5$ & 1.1 \\  
C &  5 & 1.3& 2 & $5 \times 10^4$ & 7.5  \\  
\enddata
\tablecomments{All models assume $R = 10^{13}$ cm,
$t_\mathrm{inj} = 8 R/c$, and $t_\mathrm{esc} = 5 R/c$.
The increase of $\gamma_\mathrm{max}$ during electron injection time is assumed
and the values of $\gamma_\mathrm{max}$ listed above are those at $t = t_\mathrm{inj}$.
}
\end{deluxetable}

\end{document}